\documentclass{PoS}

\title{Longitudinal Double-Spin Asymmetries for Dijet Production at Intermediate Pseudorapidity in Polarized $pp$ Collisions at $\sqrt{s}$ = 200 GeV}

\ShortTitle{}

\author{\speaker{Ting Lin for the STAR Collaboration}\\
        Texas A\&M University\\
        E-mail: \email{tinglin.physics@gmail.com}}

\abstract{One of the primary goals of the RHIC spin program is to determine the spin-dependent gluon distribution, $\Delta g(x)$, of the proton. The measurements of the 2009 longitudinal double-helicity asymmetry, $A_{LL}$, for mid-rapidity inclusive jet and $\pi^{0}$ production place strong constraints on $\Delta g(x)$ and, for the first time, find evidence for non-zero gluon polarization values for partonic momentum fraction $x$ greater than 0.05. In contrast to inclusive jets, dijet correlation measurements provide access to partonic kinematics at leading order, and thus give better constraints on the behavior of $\Delta g(x)$ as a function of gluon momentum fraction. Furthermore, dijet measurements at higher rapidity probe the lower $x$ values where $\Delta G$ is poorly constrained.\\
In these proceedings, we present the first measurement of $A_{LL}$ for dijets with at least one jet reconstructed within the pseudorapidity range 0.8 $< \eta <$ 1.8 at STAR. The dijets were measured in polarized proton+proton collisions at a center-of-mass energy $\sqrt{s}$ = 200 GeV. Values of $A_{LL}$ are determined for several distinct event topologies, defined by the jet pseudorapidities, and span a range of parton momentum fraction $x$ down to $x \sim 0.01$. The measured asymmetries are found to be consistent with the predictions of global analyses that incorporate the results of previous RHIC measurements. They will provide new constraints on $\Delta g(x)$ in this poorly constrained region when included in future global analyses.}

\FullConference{23rd International Spin Physics Symposium - SPIN2018 -\\
		10-14 September, 2018\\
		Ferrara, Italy}

\begin{document}

\section{Introduction}
Our understanding of the proton spin structure has developed over the past decades. The proton spin can be decomposed into the contributions from the quark intrinsic spin, gluon spin and their orbital angular momenta. Deep inelastic scattering (DIS) measurements have found that the spin of the quarks accounts for only about $30\%$ of the total spin of the proton; the rest must come from the gluon spin or orbital angular momenta of the partons (\cite{deFlorian:2014yva, Nocera:2014gqa} and references therein).\par
The STAR spin program has played a very important role in unraveling the gluon's polarization inside the proton. When including the STAR 2009 inclusive jet results \cite{Adamczyk:2015}, the global analyses by the DSSV \cite{deFlorian:2014yva} and NNPDF \cite{Nocera:2014gqa} collaborations show, for the first time, a positive gluon polarization in the region of sensitivity. However in the low momentum fraction region, the gluon polarization is still poorly constrained. STAR has published several new results recently (\cite{Adamczyk:2016okk, PhysRevD.98.032011, PhysRevD.98.032013}). These new data will extend our reach in $x$ using pion and jet results at forward rapidities, and also using higher collision energies. See the talk by Christopher Dilks for an overview of these recent results \cite{CDilksSpin2018}. \par
\section{Gluon Polarization at RHIC}
\begin{figure}
\centering
  \includegraphics[width=0.6\columnwidth]{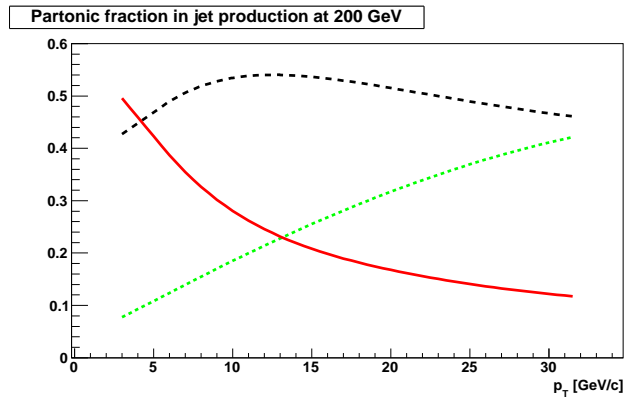}
  \caption{Partonic fraction in jet production at $\sqrt{s}$ = 200 GeV with $-0.8 \le \eta \le 1.8$ \cite{PhysRevD.86.094009, Pumplin:2002vw}, the red solid curve is is the gluon-gluon interaction, the black dashed one is the quark-gluon interaction while the green dotted line is the quark-quark interaction.}
  \label{fig:jet_partfrac}
  \vspace*{0.4cm}
\end{figure}

The longitudinal double spin asymmetry, $A_{LL}$, is the observable used to explore the gluon polarization in this analysis. $A_{LL}$ is defined as the difference of two polarized cross sections (same minus opposite helicities) over the unpolarized one, and is roughly equal to 
\begin{equation}
A_{LL} \sim \frac{\Delta f_{a}\Delta f_{b}}{f_{a}f_{b}}\hat{a}_{LL}
\end{equation}
where the $\Delta f_{a,b}$ are the helicity distributions of the two interacting partons. The $f_{a,b}$ are the unpolarized parton distribution functions, which are well constrained from other experiments, while the $\hat{a}_{LL}$ is the partonic asymmetry which can be calculated from perturbative QCD and is very large in leading order. For most RHIC kinematics, gluon-gluon and quark-gluon interactions dominate, as shown in Fig.~\ref{fig:jet_partfrac}. This makes the $A_{LL}$ measurements for jets sensitive to the gluon polarization.\par
The Relativistic Heavy Ion Collider (RHIC) \cite{Alekseev:2003sk} is located at Brookhaven National Laboratory on Long Island. It has the capability to accelerate many particle species to a wide range of energies and is the world's first and only accelerator capable of colliding polarized protons. The Solenoidal Tracker at RHIC (STAR) \cite{Ackermann:2002ad} is a large solid angle detector with charged particle tracking and electromagnetic calorimetry. The tracking is accomplished with a Time Projection Chamber (TPC) \cite{Anderson:2003ur} with full azimuthal coverage over $|\eta| < 1.3$ while the electromagnetic calorimetry is provided by Barrel Electromagnetic Calorimeter (BEMC) \& Endcap Electromagnetic Calorimeter (EEMC) and extends from $-1 \le \eta \le 2$ and $2\pi$ in azimuthal angle \cite{Beddo:2002zx, Allgower:2002zy}.\par
This analysis uses the data collected by the STAR collaboration during the 2009 running period from the longitudinally polarized proton+proton collisions at $\sqrt{s} = 200$ GeV with an integrated luminosity of 21 $\mathrm{pb^{-1}}$ and average polarization $56\%$. The jet reconstruction procedures follow those of the mid-rapidity inclusive jet and dijet analyses \cite{Adamczyk:2015, Adamczyk:2016okk}, which use the same dataset taken in 2009. Jets were reconstructed using the anti-$k_{T}$ algorithm \cite{Cacciari:2008gp}, which is implemented in the FastJet package \cite{Cacciari:2011ma}, with resolution parameter R = 0.6. Underlying event activity, which is due to the soft processes involving the beam remnants, was also considered in this analysis and was subtracted from the measured jets using the off-axis cone method \cite{ALICE:2014dla, zchang2016}.\par
Compared to the inclusive jet probes, dijets can capture more information of the hard scattering and provide a direct link to the initial kinematics, so they may place better constraints on the functional form of the gluon polarization. The dijet selection procedure in this analysis also follows those used in the STAR 2009 mid-rapidity dijet measurements \cite{Adamczyk:2016okk} but with a much wider pseudorapidity range. For details, see \cite{PhysRevD.98.032011}.\par
\section{Methods and Results}
The tracking efficiency for the STAR TPC is $\sim85\%$ for $|\eta| \le 1.0$ and then falls to about $50\%$ at $|\eta| \sim 1.3$. This falling tracking efficiency in the intermediate pseudorapidity region makes the jet reconstruction a challenge in the EEMC region. The reconstructed jets in the Endcap region will have lower $p_{T}$ and mass with fewer tracks, and will skew the extraction of the initial partonic momenta. A machine learning regression method was used to correct the measured jet $p_{T}$ and mass on a jet-by-jet basis \cite{PhysRevD.98.032011}. The final dijet invariant masses are calculated using the corrected jet transverse momenta and masses from machine learning.\par
\begin{figure*}
\centering
\begin{minipage}{.49\textwidth}
    \includegraphics[width=1\columnwidth]{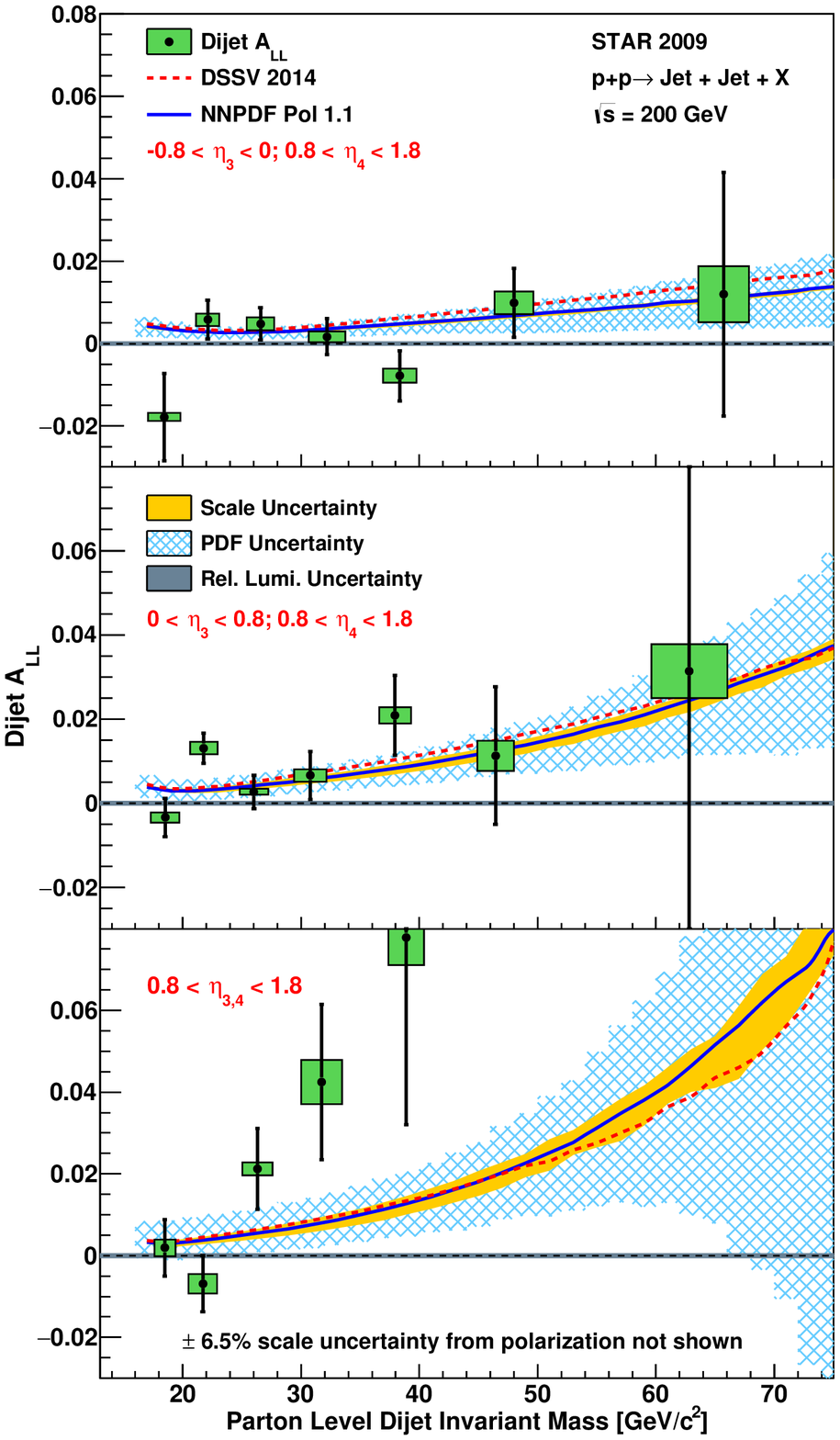}
    \caption{$A_{LL}$ as a function of parton-level invariant mass for dijets with the East Barrel-Endcap (top), West Barrel-Endcap (middle) and Endcap-Endcap (bottom) event topologies. Figure taken from \cite{PhysRevD.98.032011}.}
    \label{fig:ALL_DiffTopo}
\end{minipage}
\hfill
\begin{minipage}{.49\textwidth}
  \includegraphics[width=1\columnwidth]{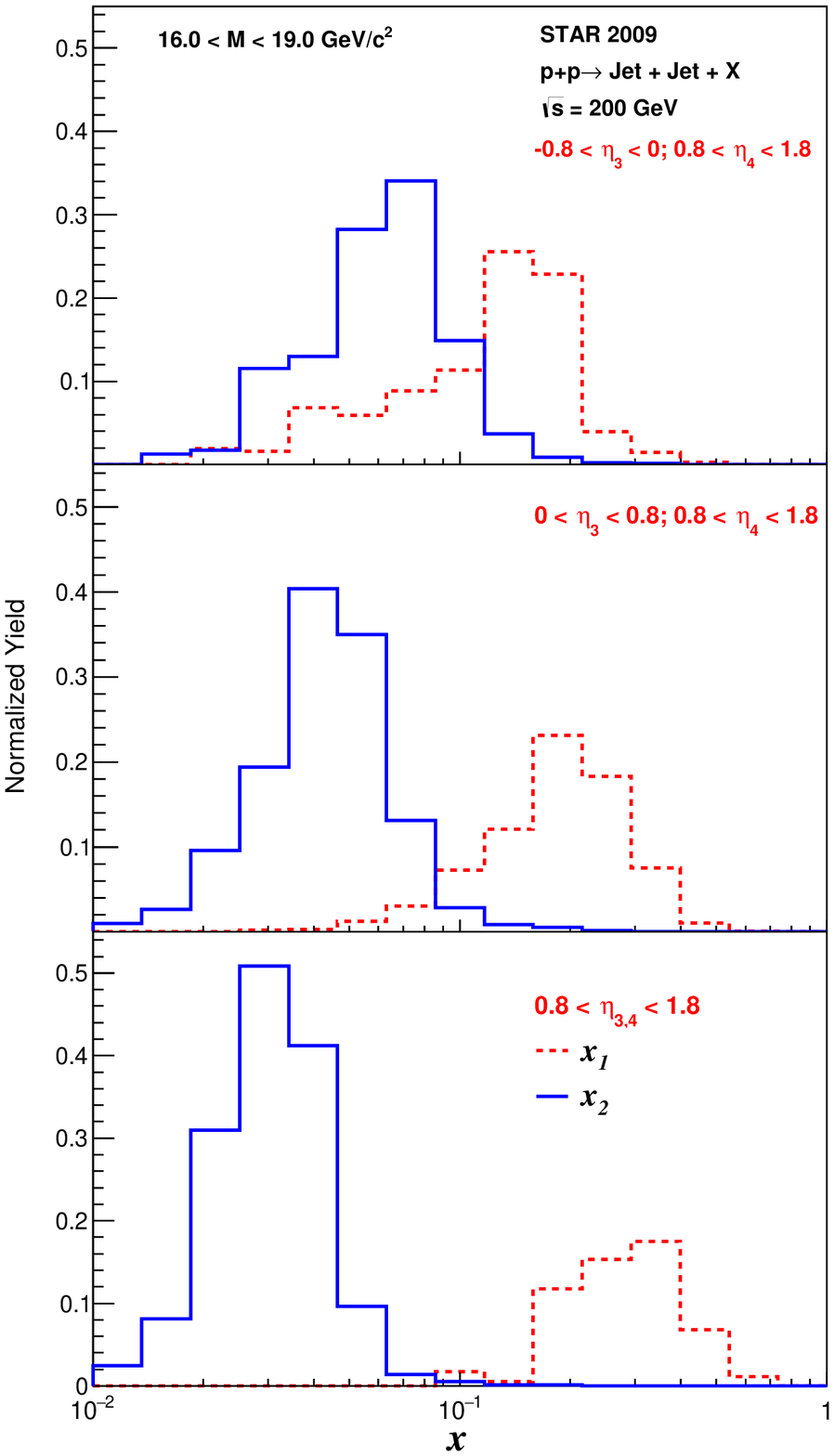}
  \caption{The distributions of the parton $x_1$ and $x_2$, which have been weighted by the partonic $\hat{a}_{LL}$, from \textsc{Pythia} detector level simulations at $\sqrt{s}$ = 200 GeV for different jet pseudorapidity ranges. Figure taken from \cite{PhysRevD.98.032011}.}
  \label{fig:Dijet_Kinematics}
\end{minipage}
\end{figure*}

The final results for the dijet longitudinal double spin asymmetries are plotted as a function of dijet invariant mass as shown in Fig.~\ref{fig:ALL_DiffTopo}. These results are separated into three dijet event topologies: dijets in which one jet is detected in the east half of the Barrel EMC ($-0.8 < \eta_{\rm jet} < 0.0$) or in the west half of the Barrel EMC ($0.0 < \eta_{\rm jet} < 0.8$), while the other is in the Endcap ($0.8 < \eta_{\rm jet} < 1.8$); and events in which both jets fall in the Endcap. The $A_{LL}$ asymmetry results presented in these figures are compared to the theoretical model predictions from the DSSV and NNPDF groups. These two theory curves were generated using the DSSV2014 \cite{deFlorian:2014yva} and NNPDFpol1.1 \cite{Nocera:2014gqa} polarized PDF sets, which also include the data from the STAR 2009 inclusive jet analysis \cite{Adamczyk:2015}.\par
The momentum fractions, $x_{1}$ and $x_{2}$, carried by the two interacting partons in the hard scattering, which can be probed by these results, are shown in Fig.~\ref{fig:Dijet_Kinematics} for the lowest dijet mass bin. The $x_{1}$ is always associated with the beam heading towards the higher pseudorapidity region. The momentum fraction distributions are also separated into three different event topologies and map to the asymmetry results. The various event topologies probe different ranges of the momentum fractions, and as jets go to higher pseudorapidity, $x_{2}$ shifts to lower values and the separation between the $x_{1}$ and $x_{2}$ increases.\par
\begin{figure}
\centering
  \includegraphics[width=0.5\columnwidth]{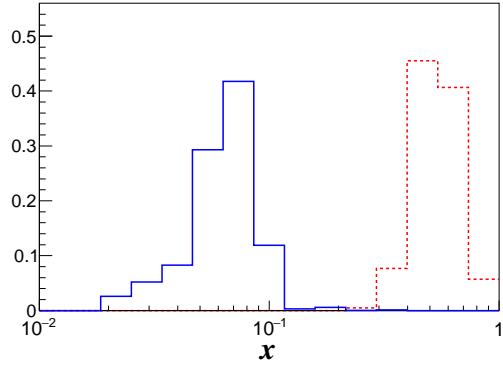}
  \caption{The distributions of the parton $x_1$ (red, dashed) and $x_2$ (blue, solid) for $0.8 \le \eta_{3,4} \le 1.8$ at dijet invariant mass $34 \le M \le 41 GeV/c^{2}$.}
  \label{fig:ALL_partfrac}
  \vspace*{0.4cm}
\end{figure}
\begin{figure}
\centering
  \includegraphics[width=0.5\columnwidth]{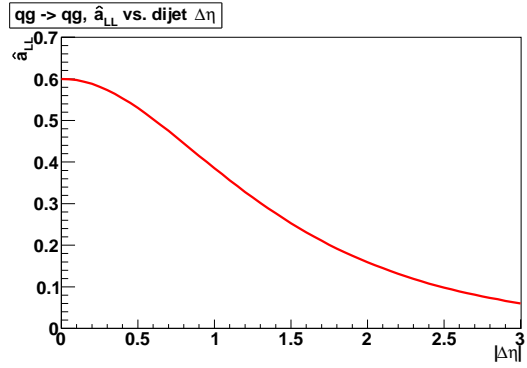}
  \caption{Dijet partonic $\hat{a}_{LL}$ as function of dijet $\Delta \eta$.}
  \label{fig:Dijet_partall}
  \vspace*{0.4cm}
\end{figure}

The smallest momentum fraction region ($x_{2} \sim 0.03$) that can be probed by these results is the first data point measured at low dijet mass values in the forward-forward dijet topology where both jets are going to the higher pseudorapidity range, as shown in the bottom panel of Fig.~\ref{fig:Dijet_Kinematics}. In contrast, the data point measured at higher dijet invariant mass at the same dijet topology is sensitive to the larger $x$ values ($x_{1} \sim 0.5$) as can be seen in Fig.~\ref{fig:ALL_partfrac}. Coupled with the knowledge from the unpolarized parton distribution functions, the low momentum fractions $x_{2}$ are always dominated by gluons while $x_{1}$ are most often valence quarks. The kinematic region of the forward-forward dijet topology maximizes the partonic asymmetry; this feature can be seen from the partonic asymmetry as a function of dijet $\Delta \eta$ curve in Fig.~\ref{fig:Dijet_partall}. Future STAR $pp$ 200 GeV data will improve the measurements in both low and high $x$ regions.\par

\section{Conclusion}
We have presented the first measurement of the longitudinal double spin asymmetry $A_{LL}$ for dijets at intermediate pseudorapidity ($0.8 \le \eta \le 1.8$). The results are in good agreement with recent theory predictions and should help to reduce the global analysis uncertainties at lower momentum fraction $x$. With the increased statistics from 2012, 2013 and 2015, new STAR data will help to further constrain the value and shape of the gluon polarization in the proton. This work was supported in part by the U.S. Department of Energy Grant DE-SC0017982.\par

\end{document}